\documentclass[onecolumn,showpacs,floatfix,aps]{revtex4}
\usepackage{amssymb,amsmath}
\usepackage{subfigure}
\usepackage[dvips]{graphics,color}
\usepackage{epsfig}\usepackage{float}

\newcommand{\hko}{\hookrightarrow}

\newcommand{\n}{\nonumber\\}

\newcommand{\ih}{\'{\i}}
\newcommand{\clp}{\cl_{p,q}}

\newcommand{\vcz}{\overset{\zeta}}

\newcommand{\med}{\medbreak}

\newcommand{\fn}{\footnotesize}

\newcommand{\cv}{\circ}
\newcommand{\di}{\diamond}

\newcommand{\bec}{\begin{center}}
\newcommand{\eec}{\end{center}}

\newcommand{\bea}{\begin{array}}
\newcommand{\ear}{\end{array}}

\newcommand{\bfr}{\begin{flushright}}

\newcommand{\efr}{\end{flushright}}
\newcommand{\noi}{\noindent}\newcommand{\Ra}{\rightarrow}
\newcommand{\ri}{\rightarrow}
\newcommand{\me}{\frac{1}{2}}

\newcommand{\cl}{{\mt{C}}\ell}

\newcommand{\RR}{\mathbb{R}}\newcommand{\op}{\oplus}
\newcommand{\HH}{\mathbb{H}}

\newcommand{\be}{\beta}
\newcommand{\ot}{\otimes}
\newcommand{\la}{\Lambda}
\newcommand{\bege}{\begin{equation}}
\newcommand{\enge}{\end{equation}}
\newcommand{\w}{\wedge}
\newcommand{\g}{\gamma}

\newcommand{\si}{\sigma}

\newcommand{\beq}{\begin{eqnarray}}\newcommand{\benu}{\begin{enumerate}}\newcommand{\enu}{\end{enumerate}}
\newcommand{\eeq}{\end{eqnarray}}
\newcommand{\mt}{\mathcal}

\newcommand{\vv}{{\bf v}}
\newcommand{\ee}{{\bf e}}
\newcommand{\uu}{{\bf u}}

\newcommand{\ww}{{\bf w}}
\newcommand{\CC}{\mathbb{C}}

\newcommand{\lam}{\lambda}
\newcommand{\non}{\nonumber\\}

\newcommand{\OO}{\mathbb{O}}
\newcommand{\dl}{{\bf{\delta}}}

\newcommand{\im}{\mathfrak{i}}\newcommand{\jm}{\mathfrak{j}}\newcommand{\km}{\mathfrak{k}}
\newcommand{\up}{\Upsilon}

\newcommand{\mk}{\mathfrak}

\newcommand{\clt}{{\mt{C}}\ell_{3,0}}
\newcommand{\cle}{{\mt{C}}\ell_{1,3}}

\newcommand{\bx}{\begin{pmatrix}}
\newcommand{\ex}{\end{pmatrix}}
\newcommand{\vcx}{\varepsilon}

\usepackage{amsfonts}

\begin{document}

\title{Isotopic liftings of Clifford algebras and applications in elementary particle mass matrices}

\author{R. da Rocha}
\email{roldao.rocha@ufabc.edu.br, roldao@ifi.unicamp.br}
\affiliation{Centro de Matem\'atica, Computa\c c\~ao e Cogni\c c\~ao\\
Universidade Federal do ABC, 09210-170 Santo Andr\'e, SP, Brazil
\\and\\
Instituto de F\'{\i}sica "`Gleb Wataghin"', Universidade Estadual de Campinas, Unicamp,
13083-970 Campinas, SP, Brazil}
\author{J. Vaz, Jr.}
\affiliation{Departamento de Matem\'atica Aplicada, IMECC, Unicamp, CP 6065, 13083-859, Campinas, SP, Brazil. }
\email{vaz@ime.unicamp.br}


\begin{abstract}

Isotopic liftings of algebraic structures are investigated in the context of Clifford algebras, 
where it is defined a new product involving an arbitrary, but fixed, element of the Clifford algebra.
This element acts as the unit with respect to the introduced product, and is called \emph{isounit}. We construct isotopies in both associative 
and non-associative arbitrary algebras, and examples of these constructions are exhibited 
using Clifford algebras, which although  associative, can generate the octonionic, non-associative, algebra. 
The whole formalism is developed in a Clifford algebraic arena, giving also the necessary 
pre-requisites to introduce isotopies of the exterior algebra. 
The flavor hadronic symmetry of the six $u,d,s,c,b,t$ quarks is shown to be \emph{exact}, when 
the generators of the \emph{isotopic} Lie algebra $\mk{su}(6)$ are constructed, and the unit of the isotopic Clifford algebra
is shown to be a function of the six quark masses. The limits constraining the parameters, that are entries of the 
representation of the isounit in the isotopic group SU(6),  are based on the most recent limits imposed on quark masses.
  \end{abstract}
\maketitle


\section{Introduction} 

Some limitations concerning the description of physical theories, owning non-canonical, non-unitary and non-lagrangian character, 
have  motivated  investigations about a wider class of formalisms used to describe such theories, 
the so-called isotopies of mathematical structures. The isotopic lifting of such structures
allows the physical theories to be described in a straightforward canonical, unitary and Lagrangian formalism \cite{kad1, kad2, san21,san223,san23,654,san01,san02,san05,san2},  
by maps from Lagrangian, linear and local theories to more general ones, envolving a 
non-linear, non-local and non-Lagrangian character. These later are led to the former when formulated in an isospace, 
 endowed with a new product in the context of the Clifford algebras, 
with respect to which the unit is now a fixed, but arbitrary, element $\zeta$ of the Clifford algebra.
The inverse of such element is called \emph{isotopic element}, and shall be used to define
the new product that endows the \emph{Clifford isotopic algebra}, to be precisely defined in this article. 
These isotopic concepts are entirely related to the $q$-deformations of algebraic structures, to which have a one-to-one correspondence
to the isotopic liftings of algebras \cite{san02}. 
 Although in e.g. \cite{san21}  isotopies of symplectic and
other geometries are included, the present paper presents for the first time the
isotopies of Clifford algebras with significant applications.

In what follows we define isotopic Clifford algebras, 
and subsequently the formalism developed is applied in some aspects of Quantum Field Theory, e.g., 
the description of the flavor SU(6) symmetry as an \emph{exact} symmetry 
among the six quarks, if they are to be viewed as components of an element
of the carrier representation space of the isotopic group SU(6)$_\zeta$ associated with the group SU(6), in the context 
of the isotopic Clifford algebra $\cl_{12,0}$.
As a consequence, all six quarks must have the same mass in \emph{isospace}, which brings an immediate constraint 
among the elements that constitute the matrix representing the Santilli's isounit, here emulated in a Clifford algebraic context. 
The isounit is shown to be a function of
quark masses, whose original values are retrieved when an eigenvalue \emph{iso}equation, or equivalently, 
the expected value defined in isospace, is used. 
The isotopic Lie algebra $\mk{su}(6)_\zeta$, associated with the Lie algebra $\mk{su}(6)$,
is constructed in the context of the isotopic lifting of the Clifford algebra $\cl_{12,0}$. 
More generally, the isotopic lifting of $\mk{su}$($n$) is described in the context of the 
isotopic lifting of the Clifford algebra $\cl_{2n,0}$, emulating a similar construction in \cite{romo}.

We illustrate the general method to be used, by firstly describing the flavor symmetry among the $u,d$ and
  $s$ quarks as an exact symmetry
of the isotopic SU(3)$_\zeta$ group, constructed via the isotopic lifting of the Dirac-Clifford algebra $\cle(\CC)=\CC\ot\cle$.
In this context the isotopic group SU(3)$^f_\zeta \times$ SU(2)$_\zeta \times$ U(1)$_\zeta$ is obtained using solely
the isotopic lifting of $\cle(\CC)$. Here SU(3)$^f$ denotes the flavor group SU(3) and has nothing to do 
with the SU(3) gauge group associated with strong interactions. Hereon we omit the index ${}^f$ and denote SU(3)$^f$ solely by SU(3).
We emphasize that the isotopies of SU(3) and the proof of their local isomorphism to the
conventional SU(3) symmetry were first proved in \cite{san05} and papers
quoted therein. 
After introducing the iso-Gell-Mann matrices, as particular cases of the most general representation in the isotopic $\mk{su}(3)$ Lie 
algebra,
analogously to \cite{san22,san18,animalu},  
the behavior of some quantum operators acting on the carrier fundamental representation space of the isotopic  SU(3) group is investigated.

In terms of its applications, 
the main aim of this paper is to obtain an exact flavor symmetry encompassing all the six quarks via the isotopic lifting 
of the generators of the group SU(6). The parameters that define the isotopy are shown to be functions of the quark masses, and are delimited by the most recent limits of quark masses.

This article is organized as follows: in Sec. II a brief review on Clifford algebras is presented, and after discussing associative 
isotopies in Sec. III, in Sec. IV the  isotopic liftings of non-associative algebras is presented. In Sec. V $\zeta$-fields are presented, and in 
Sec. VI we investigate the so-called 
Clifford admissible products in the context of the $\zeta$-applications. In  Sec. VII the isotopic lifting of exterior algebras 
is introduced via Clifford isotopic algebras and in Sec. VIII and Sec. IX a complete formulation concerning the isotopic 
lifting of spacetime algebra is presented in order to introduce the heterodimensional isotopic lifting of the group SU(3).
In Sec. X the more general case describing the isotopic generators of the Lie group SU($n$) is constructed, in the light of the 
corresponding standard construction \cite{romo}. 
Finally in Sec. XI, applications to QFT are presented and we show 
how to suitably construct an isotopy in such a way that in isospace the six quarks 
have equal masses, and consequently the SU(6) flavor symmetry becomes an exact symmetry in isospace.
In the Appendix the isotopic lifting of SU(6) is presented via the isotopic lifting of the Clifford algebra $\cl_{12,0}$.

\section{Preliminaries}
Let $V$ be a finite $n$-dimensional real vector space and $V^*$ denotes its dual. 
We consider the tensor algebra $\bigoplus_{i=0}^\infty T^i(V)$ from which we
restrict our attention to the space $\Lambda(V) = \bigoplus_{k=0}^n\Lambda^k(V)$ of multivectors over $V$. $\Lambda^k(V)$
denotes the space of the antisymmetric
 $k$-tensors, isomorphic to the  $k$-forms vector space.  Given $\psi\in\Lambda(V)$, $\tilde\psi$ denotes the \emph{reversion}, 
 an algebra antiautomorphism
 given by $\tilde{\psi} = (-1)^{[k/2]}\psi$ ([$k$] denotes the integer part of $k$). $\hat\psi$ denotes 
the \emph{main automorphism or graded involution},  given by 
$\hat{\psi} = (-1)^k \psi$. The \emph{conjugation} is defined as the reversion followed by the main automorphism.
  If $V$ is endowed with a non-degenerate, symmetric, bilinear map $g: V^*\times V^* \rightarrow \RR$, it is 
possible to extend $g$ to $\la(V)$. Given $\psi=\uu^1\w\cdots\w \uu^k$ and $\phi=\vv^1\w\cdots\w \vv^l$, for $\uu^i, \vv^j\in V^*$, one defines $g(\psi,\phi)
 = \det(g(\uu^i,\vv^j))$ if $k=l$ and $g(\psi,\phi)=0$ if $k\neq l$. The projection of a multivector $\psi= \psi_0 + \psi_1 + \cdots + \psi_n$,
 $\psi_k \in \la^k(V)$, on its $p$-vector part is given by $\langle\psi\rangle_p$ = $\psi_p$. 
 Given $\psi,\phi,\xi\in\Lambda(V)$, the  {\it left contraction} is defined implicitly by 
$g(\psi\lrcorner\phi,\xi)=g(\phi,\tilde\psi\w\xi)$. 
 For $a \in \RR$, it follows that 
 ${\bf v} \lrcorner a = 0$. Given $\vv\in V$, the Leibniz rule 
${\bf v}\lrcorner (\psi \w \phi) = ({\bf v} \lrcorner \psi) \w \phi + \hat\psi \w ({\bf v} \lrcorner \phi)$ holds. The 
 {\it right contraction} is analogously defined 
$g(\psi\llcorner\phi,\xi)=g(\phi,\psi\w\tilde\xi)$
 and its associated Leibniz rule $
(\psi \w \phi) \llcorner {\bf v} = \psi \w (\phi \llcorner {\bf v}) + (\psi \llcorner {\bf v}) \w \hat\phi
$ holds. Both contractions are related by 
${\bf v} \lrcorner \psi = -\hat\psi \llcorner {\bf v}$.
The Clifford product between $\ww\in V$ and $\psi\in\la(V)$ is given by $\ww\psi = \ww\w \psi + \ww\lrcorner \psi$.
 The Grassmann algebra $(\la(V),g)$ 
endowed with the Clifford  product is denoted by $\cl(V,g)$ or $\cl_{p,q}$, the Clifford algebra associated with $V\simeq \RR^{p,q},\; p + q = n$.
In what follows $\RR,\CC$ and $\HH$ denote respectively the real, complex and quaternionic  fields.
\section{Associative isotopic algebras}
Consider a $\CC$-associative algebra ${\mt A}$ endowed with a product  $AB$ denoted by juxtaposition, where $A, B\in{\mt A}$, 
and let $\zeta\in{\mt A}$ be a fixed, but arbitrary element of ${\mt A}$. 
The product $\di: {\mt A}\times {\mt A}\ri {\mt A}$ is given by  
\bege\label{75}
{}{A\di B:= A\zeta^{-1}B = (A\zeta^{-1})B = A(\zeta^{-1}B)}.
\enge\noi Clearly  $\zeta$ is the unit of ${\mt A}$ with respect to the $\di$-product, since  
$A\di \zeta = \zeta\di A = A$, for all  $A\in {\mt A}$. Since $\zeta$ is assumed to be always invertible, 
the product $\di$ is not automorphic to the product of the original algebras \cite{ghy}.

Given  $A \in{\mt A}$, $\zeta$-applications are defined as
\bege  \label{u}
{\di\, A := \zeta^{-1}A},\qquad\quad 
{\vcz{A}:= A \zeta}\enge\noi where the juxtaposition denotes the product in ${\mt A}$. 
 All the formalism to be developed hereon is motivated by definitions in Eqs.(\ref{u}). 

 The \emph{isotope}-$\zeta$ of the algebra ${\mt A}$, denoted by ${\mt A}_\zeta$, is defined as being 
the underlying vector space  of the algebra ${\mt A}$, with multiplication given by Eq.(\ref{75}). 
 The action of the isotopic algebra ${\mt A}_\zeta$ on physical states, generally described by elements of a Hilbert space ${\mt H}$ --- 
which is an ideal on which operators in ${\mt A}_\zeta$ acts on ---
comes from the definition of the isotope-$\zeta$ of an ${\mt A}$-module. Consider  $V$ a left unital ${\mt A}$-module, with respect to the composition 
 $A\vv$, where $A\in{\mt A}$, $\vv\in V$. Here $V$ must be a left ideal of ${\mt A}$.
From the map 
\beq
{\mt A}_\zeta\times V&\Ra& V\n
(A,\vv)&\mapsto& A\di \vv = A\zeta^{-1} \vv, 
\eeq\noi the ${\mt A}$-module  $V$  
becomes a left unital  ${\mt A}_\zeta$-module  $V_\zeta$, since $\zeta\di \vv = \zeta\zeta^{-1} \vv = \vv$, forall $\vv\in V$.

The product 
\beq
\di:{\mt A}\times{\mt A}&\ri&{\mt A} \n
(A,B)&\mapsto& A\di B
\eeq\noi can also be extended in order to encompass elements $\vcz{A}, \vcz{B}\in{\mt A}_\zeta$. 
Indeed, given $\vcz{A},\vcz{B}\in {\mt A}_\zeta$, it is immediate that  
\beq
\vcz{A}\di \vcz{B} = A\zeta\zeta^{-1}B\zeta = AB\zeta \in {\mt A}_\zeta, 
\eeq\noi i.e., with respect to the product $\di$, the elements $\vcz{A}, \vcz{B}$ inherit the structure of the product $AB$ in ${\mt A}$. 
This concept shall be useful in order to define exterior algebras isotopic liftings in Sec. \ref{isoae}.

\section{Non-associative isotopy}

In this case the algebra ${\mt A}$ is a non-associative $\CC$-algebra, and therefore the last equality in Eq.(\ref{75}) 
does not hold anymore. Given  $\zeta\in{\mt A}$ fixed, 
but arbitrary, the non-associative {isotope-$\zeta$} of ${\mt A}$, denoted by ${\mt A}_{(\zeta)}$, is defined by the 
 multiplication
\bege\label{76}
{}{A\di_\zeta B:= A(\zeta^{-1}B)} 
\enge\noi while the  \emph{$\zeta$-isotope} of ${\mt A}$, denoted by $_{(\zeta)}{\mt A}$, is defined by the relation
\bege\label{77}
{}{A_\zeta\negmedspace\di B:= (A\zeta^{-1})B} 
\enge\noi
We verify that, while  $\zeta$ is the right unit of the algebra ${\mt A}_{(\zeta)}$ with respect to the product given by Eq.(\ref{76}),
$\zeta$ is also the left unit of $_{(\zeta)}{\mt A}$ with respect to the product given by Eq.(\ref{77}).
The product $\di_\zeta$ defines uniquely the  isotope-$\zeta$  ${\mt A}_{(\zeta)}$ of ${\mt A}$, while in a similar way the product 
$_\zeta\di$ defines the $\zeta$-isotope  $_{(\zeta)}{\mt A}$ of ${\mt A}$. 
Naturally the product $\di_\zeta$ can be extended to elements $\vcz{A}, \vcz{B}$ in the isotope-$\zeta$  ${\mt A}_{(\zeta)}$ of ${\mt A}$, 
 in such way that for this non-associative case it follows that 
\beq
\vcz{A}\di_\zeta \vcz{B} &:=& \vcz{A}(\zeta^{-1} \vcz{B})\n
&=& (A\zeta) (\zeta^{-1} B\zeta)\label{zent}
\eeq\noi In this way it is then possible to define the product 
 $A\cv_{\zeta}B := (A\zeta)(\zeta^{-1}B)$ 
from Eq.(\ref{zent}), which extends the $X$-product introduced in the octonionic algebra $\OO$ context, to any non-associative algebra ${\mt A}$.
The $X$-product was originally introduced  in order to correctly define 
the transformation rules for bosonic (vector) and fermionic (spinor) fields 
on the tangent bundle over the 7-sphere $S^7$ \cite{ced}. This product is also
closely related to the parallel transport of sections of the tangent bundle, at $X\in S^7$, i.e., $X\in\OO$ such that $\bar{X}X = X\bar{X} = 1$. 
The $X$-product is also shown to be twice 
the parallelizing torsion \cite{mart}, given by the torsion tensor, and in particular, it is used to investigate the 
 $S^7$ Ka$\check{{\rm c}}$-Moody algebra \cite{ced,kac} and to
obtain triality maps and $G_2$ actions \cite{dix,beng}. Also,  
it leads naturally to remarkable geometric and topological properties, for instance the Hopf fibrations   
$S^3\cdots S^7 \Ra S^4$ and $S^8\cdots S^{15} \Ra S^7$ \cite{abg}, and twistor formalism in ten dimensions \cite{mart,rocha}.
Generalizations of these topics are developed in \cite{ghy}.
We also extend the product  
$_\zeta\di$ to the $\zeta$-isotope  $_{(\zeta)}{\mt A}$ of ${\mt A}$, 
in such a way that for this case we have
\beq
\vcz{A}_\zeta\negmedspace\di \vcz{B} &:=& (\vcz{A}\zeta^{-1}) \vcz{B}\n&=& (A\zeta\zeta^{-1})(B\zeta)\n
&=& A (B\zeta).
\eeq

The definitions of the left unital ${\mt A}_{(\zeta)}$-module and  the $_{(\zeta)}{\mt A}$-module for the cases given by Eqs.(\ref{76}, \ref{77}) 
follow naturally from their respective definitions.
\med
{\bf Example 1}: 
 The octonion algebra $\OO$ can be generated by a basis  $\{\ee_0 = 1,\ee_a\}_{a=1}^7$
  in the underlying paravector space \cite{loun,bay2} $\RR\op\RR^{0,7}\hko\cl_{0,7}$, endowed with the standard 
octonionic product $\cv:\OO\times\OO\Ra\OO$, which can be constructed using the Clifford algebras $\cl_{0,7}$  as
\bege
A\cv B = \langle AB(1 - \psi)\rangle_{0\oplus 1},\quad A,B\in\RR\op\RR^{0,7},
\enge\noi where $\psi = \ee_1\ee_2\ee_4 + \ee_2\ee_3\ee_5 + \ee_3\ee_4\ee_6 + \ee_4\ee_5\ee_7 + \ee_5\ee_6\ee_1 + \ee_6\ee_7\ee_2 + \ee_7\ee_1\ee_3 \in \Lambda^3(\RR^{0,7})\hko \cl_{0,7}$ and 
the juxtaposition denotes the Clifford product \cite{loun}. The idea of introducing the octonionic product from the Clifford product in this context is to 
present hereon in this example our formalism using a  Clifford algebraic arena. 
It is now immediate to verify the usual rules between basis elements under the octonionic product:
\bege\label{taboc}
\ee_a\cv \ee_b = \vcx_{ab}^c \ee_c - \delta_{ab} \quad (a,b,c =1,\ldots,7), 
\enge\noi where we denote  $\vcx_{ab}^c = 1$
for the cyclic permutations ($abc$) = (124),(235),(346),(457),(561),(672) and (713).  All the relations above can be expressed as $\ee_a\cv \ee_{a+1} = \ee_{a+3\mod 7}$.

Now, defining  $\zeta = \ee_1$, the isotope-$\zeta$ $\OO_{(\zeta)}$ related to the octonionic algebra $\OO$, is given by  
the multiplication 
\bege
A\di_\zeta B= A\cv (\ee_1^{-1}\cv B),
\enge\noi and the $\zeta$-isotope $_{(\zeta)}\OO$ of $\OO$ is defined by 
\bege
A\;_\zeta\negmedspace\di B = (A\cv \ee_1^{-1})\cv B.
\enge\noi  For the particular cases where $A = \ee_2$ and $B = \ee_4$ it follows that
\beq
\ee_2\di_\zeta \ee_5 &=& \ee_2 \cv (\ee_1^{-1}\cv \ee_5)\n
&=& \ee_2\cv (-\ee_6)\n
&=& -\ee_7
\eeq\noi while 
\beq
\ee_2\,_\zeta\negmedspace\diamond \ee_5 &=& (\ee_2 \cv \ee_1^{-1})\cv \ee_5\n
&=& \ee_4\cv \ee_5\n
&=& \ee_7.
\eeq\noi
\med

\section{$\zeta$-fields and isocomplex fields}
An isotopy of the unit $1\in {\mt A}$ is defined to be the map $1\mapsto \zeta  = \zeta(x)$. 
For consistency of the formalism, the associative products between operators are led to their corresponding isotopic (associative) partners: 
\bege\label{78}
AB\mapsto A\di B = A \zeta^{-1}B,\qquad \text{$\zeta$ fixed}.
\enge\noi As we have just seen, the element  $\zeta$ is the unit with respect to the product 
 $\di$, also denominated  \emph{isounit}. On the other hand $\zeta^{-1}$ is called \emph{isotopic element}. 

The field $\CC = \CC(a, + , \times)$ with elements $a\in\CC$, ordinary sum 
 $a_1 + a_2$ and multiplication $a_1\times a_2 = a_1a_2$ 
is isotopically lifted to the isofield  $\vcz{\CC}(\mk{a}, \vcz{+}, \di)$, where the isocomplex numbers  
(heretofore denoted by gothic characters) are given by $\mk{a}:= a\zeta$, the sum is expressed as  $\mk{a}_1 \vcz{+} \mk{a}_2:= (a_1 + a_2) \zeta$ and
the isomultiplication by $\mk{a}_1 \di \mk{a}_2 = \mk{a}_1 \zeta^{-1} \mk{a}_2 = (a_1 a_2) \zeta$. The fields $\CC$ and 
$\vcz{\CC}$ are shown to be isomorphic \cite{kad1}. Note that given an operator  $A\in {\mt A}$, the isoproduct between isoscalars 
and such operator is given by  $\mk{a} \di A = a\zeta\zeta^{-1}A = aA$.

We should mention the effect that the lack of
use of Santilli's isofield activates the theorems of catastrophic
inconsistencies \cite{654} in the case when non-canonical or non-unitary theory is not formulated on
Santilli's isofields.

\section{Clifford isotopies via associative $\zeta$-product}
\label{secliei}
From this 
section on the algebra ${\mt A}$ is taken to be the Clifford algebra $\clp$ over the quadratic space  $\RR^{p,q}$.
It is well known that the Lie algebra $\mk{so}(p,q)\simeq \mk{spin}(p,q)$ is isomorphic to the algebra $(\Lambda^2(\RR^{p,q}),[\;,\;])$ ---
where  $[\;,\;])$ denotes the commutator ---  when  Spin($n$,0) $\simeq$ Spin(0,$n$) 
and Spin$_+$($n-1$,1) $\simeq$ Spin$_+$(1,$n-1$), $n > 4$ \cite{coun}. 
Then besides the product given by Eq.(\ref{75}), there is defined the isocommutator \cite{kad1, kad2, san01, san05, san2, san12} 
$[\;\,,\;\,]_\zeta$ defined by 
\bege\label{80}
{}{[{\psi_i},{\psi_j}]_\zeta:= {\psi_i}\di{\psi_j} - {\psi_j}\di{\psi_i} = {\mk{c}}^k_{ij}\di {\psi_k}}\qquad 
\enge\noi where $\psi_i, \psi_j, \psi_k$ are the generators of the isotopic lifting of $\mk{spin}(p,q)\hko\clp$, 
and $\mk{c}^k_{ij}:= {c}^k_{ij} \zeta$ are the isostructure constants of the Lie isoalgebra 
($\mk{spin}(p,q), [\;\,,\,\;]_\zeta$). Here, the $c^k_{ij}$ denote the structure constants of the Lie algebra  $\mk{spin}(p,q)$.

The product $\di:\clp\times\clp\Ra\clp$ has a structure of the Clifford product, since given $\psi,\phi\in\clp$ it follows that 
\beq
\vcz{\psi}\di\vcz{\phi} + \vcz{\phi}\di\vcz{\psi} &=& \psi \zeta\zeta^{-1}\phi \zeta + \phi \zeta\zeta^{-1}\psi \zeta\non
&=& (\psi\phi + \phi\psi)\zeta =  2g(\psi,\phi)\zeta\non
&\equiv& 2 \mk{g}(\vcz\psi, \vcz\phi), \label{acze}
\eeq\noi where $\mk{g}(\vcz\psi, \vcz\phi)\in\vcz\RR$ is defined to be the iso-metric $g(\psi,\phi)\zeta$, and
  the element $\zeta\in\cl_{p,q}$ acts as the unit with respect to the product $\di$.

Consider now that the algebra $\clp$ be endowed with the commutator $[\;\;,\;\;]$ 
given by $[\psi,\phi] = \psi\phi - \phi\psi,\;\;\forall \psi,\phi\in\clp$. 
The  \emph{Clifford isotopic algebra} 
$\clp^\zeta$ is defined as being the triple ($\clp,\di,[\;\;,\;\;]_\zeta$), where the isocommutator  
\bege\label{71}
[\psi,\phi]_\zeta := \psi\di \phi - \phi\di \psi = \psi \zeta^{-1}\phi - \phi \zeta^{-1}\psi
\enge\noi can be thought as being the isotopic lifting of the commutator $[\;\,,\;\,]$. 
The Clifford algebra $\clp^\zeta$ inherits the structure of 
 $\clp$, with the difference that the relations holding in $\clp$,
with respect to the Clifford product $\psi\phi$
now are valid in the isotope  $\clp^\zeta$, with product  $\vcz\psi\di \vcz\phi$. Indeed, 
\beq
[\vcz\psi,\vcz\phi]_\zeta &=&\vcz\psi\di \vcz\phi - \vcz\phi\di \vcz\psi\n
 &=& \psi\zeta \zeta^{-1}\phi\zeta - \phi\zeta \zeta^{-1}\psi\zeta\n
&=& (\psi\phi - \phi\psi)\zeta\n
&=& [\psi,\phi]\zeta.\label{2100}
\eeq\noi 

\subsection{Clifford genotopies}

Given fixed, but arbitrary elements $\xi, \zeta\in\clp$, Eq.(\ref{71}) can be still generalized by the \emph{genocommutator}: 
\bege
[\psi,\phi]_{\zeta,\xi}:= \psi \zeta^{-1}\phi - \phi \xi^{-1}\psi,\qquad \psi,\phi\in\clp.
\enge\noi When $\xi=\zeta$ the genocommutator is led to the isocommutator given by Eq.(\ref{71}). 
Applications concerning the Clifford genotopic admissible algebras (CGAA), defined as being 
the 4-tuple $(\clp, [\;\,,\;\,]_{\zeta,\xi}, (\;\cdot\;)\xi,\; (\;\cdot\;)\zeta)$, 
can be used to investigate irreversible systems. Here $(\;\cdot\;)\xi$ and $(\;\cdot\;)\zeta$ obviously denote  right 
multiplication respectively by $\xi$ and $\zeta$.

\section{Exterior algebra isotopy}
\label{isoae}

It is well known that the exterior product and the (left) contraction can be defined in terms of the Clifford product respectively as
\bege
\vv \w \psi = \frac{1}{2}(\vv \psi + \hat\psi \vv)
\enge\noi and
\bege
\vv \lrcorner \psi = \frac{1}{2}(\vv \psi - \hat \psi \vv)
\enge\noi for all $\vv\in \RR^{p,q}, \psi\in\clp$. The most natural manner to define the exterior product isotopic lifting 
$\vv\w\psi\mapsto \vv\overset{\zeta}\w\psi$ is from Clifford algebras, via
\bege\label{zet1}
{}{\vv\overset{\zeta}\w\psi:= \me(\vv \di\psi + \hat\psi\di \vv) = \me(\vv \zeta^{-1}\psi + \hat\psi\zeta^{-1} \vv)}
\enge\noi 
In a similar style, the isotopic (left) contraction is defined as being
\bege
\vv \overset{\zeta}\lrcorner \psi = \frac{1}{2}(\vv \zeta^{-1}\psi - \hat \psi{\zeta^{-1}} \vv)
\enge\noi with an immediate analogue to the right contraction. Although these
 definitions above are correct from the formal viewpoint, we see from Eq.(\ref{acze}) that the product $\di$ (endowing $\clp^\zeta$) inherit
the structure of the original Clifford product $\psi\phi$ only whether  
computed between elements $\vcz\psi,\vcz\phi\in \vcz\clp$.
Therefore with respect to the physical applications concerning the formalism  developed, the following extensions are very useful:
\beq\label{105}
\vcz\vv \vcz\w \vcz\psi &=& \frac{1}{2}(\vcz\vv\di \vcz\psi + \hat{\vcz\psi}\di \vcz\vv)\n
&=& \frac{1}{2}(\vv \psi + \hat\psi \vv)\zeta
\eeq\noi and
\bege
\vcz\vv \vcz\lrcorner \vcz\psi = \frac{1}{2}(\vcz\vv\di \vcz\psi - \hat{\vcz\psi}\di\vcz \vv) = \frac{1}{2}(\vv \psi - \hat\psi \vv)\zeta
\enge\noi $\vv\in \RR^{p,q}, \psi\in\clp$. 
From Eq.(\ref{105}) it follows that 
\beq
\vcz\uu\vcz\w\vcz\vv &=& \me(\uu\zeta\zeta^{-1}\vv\zeta - \vv\zeta\zeta^{-1}\uu\zeta)\n
&=& (\uu\w\vv)\,\zeta,\quad\uu,\vv\in V.\label{1006}
\eeq In this way we see that the isotopic exterior product $\uu\vcz\w\vv$ indeed induces the exterior product 
$\uu\w\vv$ \emph{at the isospace}.

\section{The spacetime algebra  $\cle$}
\label{scle}
Consider an orthonormal basis $\{\ee_0, \ee_1, \ee_2, \ee_3\}$ in Minkowski spacetime $\RR^{1,3}$,  
where $\ee_\mu\ee_\nu + \ee_\nu\ee_\mu = 2\eta_{\mu\nu} = 2\, {\rm diag}\,(1,-1,-1,-1)$. An arbitrary element  $\Upsilon\in\cle$ is written as
$ \Upsilon = c + c^\mu\ee_\nu + c^{\mu\nu}\ee_{\mu\nu} + c^{\mu\nu\si}\ee_{\mu\nu\si} + h\ee_{0123},$  where $c, c^\mu, c^{\mu\nu}, h \in \RR$. We use the notation $\ee_{\mu\nu} = \ee_\mu\ee_\nu$, $\ee_{\mu\nu\rho} = \ee_\mu\ee_\nu\ee_\rho$ for $\mu\neq\nu\neq\rho$.

The 4-vector $\ee_{0123}$ is denoted by $\ee_5$ and  satisfies $(\ee_5)^2 = -1$, besides anticommutating with vectors: 
$\ee_\mu\ee_5 = -\ee_5\ee_\mu$. 
As  $\cle\simeq {\mathcal{M}}(2, \HH)$, in order to obtain a representation
 of $\cle$ in terms of matrices with quaternionic entries, the primitive idempotent $f = \me(1 + \ee_0)$ is used. A left minimal ideal of 
 $\cle$ is written as $I_{1,3} = \cle f$, which elements are written as
\bege
 \Xi = (a^1 + a^2\ee_{23} + a^3 \ee_{31} + a^4\ee_{12})f + (a^5 + a^6\ee_{23} + a^7\ee_{31} + a^8\ee_{12})\ee_5 f,
\enge
\noi where
\beq
a^1 &=& c + c^0,\quad a^2 = c^{23} + c^{023},\quad a^3 = -c^{13} - c^{013},\quad
a^4 = c^{12} + c^{012},\n a^5 &=& -c^{123} + c^{0123},\quad a^6 = c^1 - c^{01},
\quad a^7 = c^2 - c^{02},\quad a^8 = c^3 - c^{03}.
\eeq
\noi Since the equality $\ee_\mu = f\ee_\mu f + f\ee_\mu\ee_5 f - f\ee_5 \ee_\mu f -f\ee_5\ee_\mu\ee_5 f$ clearly holds,  
from the representation $\ee_\mu\in\cle$ in ${\mt M}(2,\HH)$ given by
\bege
\ee_0 = \left(\bea{cc}
 1&0\\
      0&-1\ear\right),\quad \ee_1 = 
\left(\bea{cc}0&\im\\
      \im&0\ear\right),\quad \ee_2 =
\left(\bea{cc}0&\jm\\
      \jm&0\ear\right),\quad \ee_3 =
\left(\bea{cc}0&\km\\
      \km&0\ear\right),\\ 
  \enge   
\noi where $\im,\jm,\km$ denote quaternionic units, the representations  of the ideal generators 
\bege
f = \left(\bea{cc}1&0\\
      0&0\ear\right),\quad
\ee_5 f  = \left(\bea{cc}0&0\\
      1&0\ear\right),
     \enge      \noi are obtained, and finally it is possible to write $\up\in\cle$ in ${\mt M}(2,\HH)$ as \bege\label{mak}
 {\bf{\up}} = \left(\bea{cc}
                \bea{c}
                c + c^0 + (c^{23} + c^{023})\im \\
                +(-c^{13} - c^{013})\jm + (c^{12} + c^{012})\km\\
                \quad\quad\quad\quad\quad\quad\quad\\
                (-c^{123} + c^{0123}) + (c^1 - c^{01})\im \\
                +(c^2 - c^{02})\jm + (c^3 - c^{03})\km
                \ear
                \bea{c}
                (-c^{123} - c^{0123}) + (c^1 + c^{01})\im +\\
                (c^2 + c^{02})\jm + (c^3 + c^{03})\km\\
                 \quad\quad\quad\quad\quad\quad\quad\\     
                 (c - c^0) + (c^{23} - c^{023})\im +\\
                (-c^{13} + c^{013})\jm + (c^{12} - c^{012})\km
                \ear
                \ear\right).
                					\enge
The Spin$_+$(1,3) group associated with  $\cle$ is given by 
\bege\label{sp13} 
{\rm Spin}_+(1,3) = \{R \in \cl^+_{1,3} \;|\;R\tilde{R} = 1\}
\enge 
Now taking a basis $\{e_i\}$ of Euclidean space $\RR^3$, the Clifford algebra $\clt$ over $\RR^3$
 is well known to be isomorphic to ${\mt M}(2,\CC)$ and  that the quaternionic units can be written as
$\im = e_3e_2, \jm = e_3e_1, \km = e_1e_2$. If the isomorphism $\clt\simeq {\mt M}(2,\CC)$
given by $e_i \mapsto \si_i$ is considered, where $\si_i$ denotes the Pauli matrices given by $\bx 0&1\\1&0\ex,\,\bx 0&-i\\i&0\ex\,\,\bx 1&0\\0&-1\ex$, 
  it follows that
the matrix in Eq.(\ref{mak}) can be written as 
\bege\label{angel}
\bx
a_1&b_1 & d_1 &f_1\\
-\bar{b_1}&\bar{a_1} &  -\bar{f_1}&  \bar{d_1}\\  
a_2&b_2 & d_2 &f_2\\
-\bar{b_2}&\bar{a_2} &  -\bar{f_2}&  \bar{d_2}
               \ex\enge\noi where
$a_1 = c + c^0 + i(c^{12} + c^{012}), b_1 = -c^{13} - c^{013} + i((c^{23} + c^{023}), d_1=-c^{123} + c^{0123} + i(c^3 - c^{03}), f_1=c^2 - c^{02} + i(c^1 - c^{01}),
a_2 = -c^{123} - c^{0123} + i (c^3 + c^{03}), b_2 = -c^{123} - c^{0123} - i (c^3 + c^{03}), d_2 = -c^{23} + c^{023} + i(-c^{13} + c^{013}),
 f_2 = c^{23}- c^{023} + i(-c^{13} + c^{013})$. 

\section{Isotopy $\cle^\zeta$ of $\cle$}

In this case the basis $\{{}\ee_\mu\}$ of $\RR^{1,3}$ satisfies  
 $\vcz\ee_\mu\di{}\vcz\ee_\nu + {}\vcz\ee_\nu\di{}\vcz\ee_\mu = 2\eta_{\mu\nu}\zeta$, and 
an arbitrary element 
of $\Upsilon_\zeta\in\cle^\zeta$ can be now written as
\begin{eqnarray} \Upsilon_\zeta &=& \mk{c} + \mk{c^\mu}\di{}\vcz\ee_\nu + \mk{c}^{\mu\nu}\di{}\vcz\ee_\mu\di{}\vcz\ee_\nu + \mk{c}^{\mu\nu\si}
\di{}\vcz\ee_\mu\di{}\vcz\ee_\nu\di{}\vcz\ee_\si + \mk{h}\di
{}\vcz\ee_0\di{}\vcz\ee_1\di{}\vcz\ee_2\di{}\vcz\ee_3,\nonumber
\end{eqnarray}\noi where $\mk{c}, \mk{c}^\mu, \mk{c}^{\mu\nu}, \mk{h} \in \vcz\RR$.

The isomultivector $\vcz\ee_0\di\vcz\ee_1\di\vcz\ee_2\di\vcz\ee_3$ is denoted by $\ee_5^\zeta$ and satisfies $\ee_5^\zeta\di\ee_5^\zeta = 
-\zeta$. 
Now choosing a primitive idempotent, denoted by  $f_\zeta = \me(\zeta + \vcz\ee_0)$, 
a left minimal ideal $I_{1,3}^\zeta$ associated with the isotopic algebra 
 $\cle^\zeta$ is written as  $\cle^\zeta\di f_\zeta$, which     
elements are of the form
\beq
 \Xi_\zeta &=& (\mk{a}^1 + \mk{a}^2\di{}\vcz\ee_2\di{}\vcz\ee_3 + \mk{a}^3 {}\vcz\ee_3\di{}\vcz\ee_1 + \mk{a}^4{}\vcz\ee_1\di{}\vcz\ee_2)\di f_\zeta\n\qquad\qquad&& + 
(\mk{a}^5 + \mk{a}^6\di{}\vcz\ee_2\di{}\vcz\ee_3 + \mk{a}^7\di{}\vcz\ee_3\di{}\vcz\ee_1 + \mk{a}^8\di{}\vcz\ee_1\di{}\vcz\ee_2)\di\vcz\ee_5\di f_\zeta,\qquad \mk{a}^m\in\vcz\CC,
\eeq
\noi where
\bege\begin{array}{lll}
\mk{a}^1 = \mk{c} + \mk{c}^0,&\quad&\mk{a}^2 = \mk{c}^{23} + \mk{c}^{023},\\
\mk{a}^3 = -\mk{c}^{13} - \mk{c}^{013},&\quad& \mk{a}^4 = \mk{c}^{12} + \mk{c}^{012},\\
\mk{a}^5 = -\mk{c}^{123} + \mk{c}^{0123},&\quad&\mk{a}^6 = \mk{c}^1 - \mk{c}^{01},\\
\mk{a}^7 = \mk{c}^2 - \mk{c}^{02},&\quad&\mk{a}^8 = \mk{c}^3 - \mk{c}^{03}.
\ear
\enge
Since $
\vcz\ee_\mu = f_\zeta\di\vcz\ee_\mu\di f_\zeta + f_\zeta\di\vcz\ee_\mu\di\ee_5^\zeta\di f_\zeta - f_\zeta\di\ee_5^\zeta\di\vcz\ee_\mu\di f_\zeta -f_\zeta\ee_5^\zeta\di
\vcz\ee_\mu\di\ee_5^\zeta\di f_\zeta
$ and from the representation of $\vcz\ee_\mu = \ee_\mu\zeta \in\cle^\zeta$ in ${\mt M}(2,\HH)_\zeta$ it follows that 
\beq
\vcz\ee_0 = \left(\bea{cc}
\zeta_0&\zeta_1\\
      -\zeta_2&-\zeta_3\ear\right),\quad  
\vcz\ee_1 = \bx \im\zeta_2&
\im\zeta_3\\ \im\zeta_0&\im\zeta_1\ex,\quad
\vcz\ee_2 = \bx \jm\zeta_2&
\jm\zeta_3\\ \jm\zeta_0&\jm\zeta_1\ex,\quad
\vcz\ee_3  = \bx \km\zeta_2&
\km\zeta_3\\ \km\zeta_0&\km\zeta_1\ex,
  \eeq
\noi and then it follows that
\bege
f_\zeta = \left(\bea{cc}1&0\\
      0&0\ear\right),\quad
\ee_5^\zeta\di f_\zeta  = \left(\bea{cc}0&0\\
      1&0\ear\right).
     \enge      \noi It follows that $\up\in\cle^\zeta$ is written in ${\mt M}(2,\HH)_\zeta$ as
 \bege
 {\bf{\up}}_\zeta = 
\bx
a_1&b_1 & d_1 &f_1\\
-\bar{b_1}&\bar{a_1} &  -\bar{f_1}&  \bar{d_1}\\  
a_2&b_2 & d_2 &f_2\\
-\bar{b_2}&\bar{a_2} &  -\bar{f_2}&  \bar{d_2}
               \ex\,\bx
\zeta_0&\zeta_1 & \zeta_2 &\zeta_3\\
-\bar{\zeta_1}&\bar{\zeta_0} &  -\bar{\zeta_3}&  \bar{\zeta_2}\\  
\zeta_4&\zeta_5 & \zeta_6 &\zeta_7\\
-\bar{\zeta_5}&\bar{\zeta_4} &  -\bar{\zeta_7}&  \bar{\zeta_6}
               \ex\enge\noi where the right hand side matrix is the complexification of $\zeta\in {\mt M}(2,\HH)$.
The isotopic Spin$_+^\zeta$(1,3) group associated with $\cle^\zeta$ is now defined by 
\bege\label{sp13}
{\rm Spin}^\zeta_+(1,3) = \{R \in \cl^{\zeta +}_{1,3}\,|\,R\di\tilde{R} = \zeta\}. \enge

\subsection{Heterodimensional isotopic lifting of SU(3)}
\label{su3c}
The isotopies of SU(3) and the proof of their local isomorphism to the
conventional SU(3) symmetry were first proved in \cite{san05} and papers
quoted therein.

From the examples above that show how to include the Lie algebra $\mk{su}(3)$, associated with the Lie group SU(3), in $\CC\ot\cle=\cle(\CC)$, 
it is possible to construct the isotopic lifting  $\mk{su}(3)_\zeta\hko\cle^\zeta$ as:
\subsubsection*{SU(3)$_\zeta$: case 1}
It is well known that considering an orthonormal basis 
$\{\ee^a\}$ of $\RR^{p,q}$, the relation $\ee^a\ee^b = \ee^a\w\ee^b$ holds between 
the Clifford and the exterior product. Denoting 
${\vcz{\ee^\mu}}$ by $\ee^\mu_\zeta$, we define the isotopic lifting $\mk{su}(3)_\zeta$ of $\mk{su}(3)$, 
that generates the isotopic group  SU(3)$_\zeta$, as the isotopic lifting given by
\beq\label{51}
\lambda_\zeta^1 &=& \me(\ee_\zeta^0\vcz\w\ee_\zeta^1 + i\ee_\zeta^2\vcz\w\ee_\zeta^3),\qquad \lambda_\zeta^2 = \me(\ee_\zeta^0\vcz\w\ee_\zeta^2 - i\ee_\zeta^1\vcz\w\ee_\zeta^3),\n
\lambda_\zeta^3 &=& \me(\ee_\zeta^0\vcz\w\ee_\zeta^3 + i\ee_\zeta^1\vcz\w\ee_\zeta^2),\qquad \lambda_\zeta^4 = \me(\ee_\zeta^0 + i\ee_\zeta^0\vcz\w\ee_\zeta^1\vcz\w\ee_\zeta^2,)\n
\lambda_\zeta^5 &=& \me(i\ee_\zeta^3 - \ee_\zeta^1\vcz\w\ee_\zeta^2\vcz\w\ee_\zeta^3),\qquad \lambda_\zeta^6 = \me(\ee_\zeta^0\vcz\w\ee_\zeta^2\vcz\w\ee_\zeta^3 + i\ee_\zeta^2),\n
\lambda_\zeta^7 &=& \frac{i}{2} (\ee_\zeta^1 + \ee_\zeta^0\vcz\w\ee_\zeta^1\vcz\w\ee_\zeta^3),\qquad \lambda_\zeta^8 = \frac{i}{\sqrt{3}}\ee^5_\zeta + \frac{1}{2\sqrt{3}}\ee_\zeta^0\vcz\w\ee_\zeta^3 
- \frac{i}{2\sqrt{3}}\ee_\zeta^1\vcz\w\ee_\zeta^2,\eeq where $\ee^5_\zeta:= \ee_\zeta^0\vcz\w\ee_\zeta^1\vcz\w\ee_\zeta^2\vcz\w\ee_\zeta^3$. 
\med
{\bf Example 2}: 
For instance, let us calculate, e.g., the isocommutator $[\lambda_\zeta^1, \lambda_\zeta^2]_\zeta$. Using definitions in Eq.(\ref{51})
and making use of the definitions (\ref{zet1}) and (\ref{105}) yields
\beq
 [\lambda_\zeta^1, \lambda_\zeta^2]_\zeta&=& \left[\me(\ee_\zeta^0\vcz\w\ee_\zeta^1 + i\ee_\zeta^2\vcz\w\ee_\zeta^3),
 \me(\ee_\zeta^0\vcz\w\ee_\zeta^2 - i\ee_\zeta^1\vcz\w\ee_\zeta^3)\right]_\zeta\n
&=&\frac{1}{8}[\ee_\zeta^0\di\ee_\zeta^1 - \ee_\zeta^1\di\ee_\zeta^0 + i\ee_\zeta^2\di\ee_\zeta^3 - i\ee_\zeta^3\di\ee_\zeta^2,
-i\ee_\zeta^1\di\ee_\zeta^3 +i\ee_\zeta^3\di\ee_\zeta^1 + \ee_\zeta^0\di\ee_\zeta^2 - \ee_\zeta^2\di\ee_\zeta^0]_\zeta
\eeq\noi Using Eq.(\ref{2100}) it follows that
\beq
 [\lambda_\zeta^1, \lambda_\zeta^2]_\zeta
&=&\frac{1}{8}[\ee^0\ee^1 - \ee^1\ee^0 + i\ee^2\ee^3 - i\ee^3\ee^2,
-i\ee^1\ee^3 +i\ee^3\ee^1 + \ee^0\ee^2 - \ee^2\ee^0]\zeta\n
&=& \frac{1}{4}(\ee^0\ee^1 - \ee^1\ee^0 + i\ee^2\ee^3 - i\ee^3\ee^2)(-i\ee^1\ee^3 +i\ee^3\ee^1 + \ee^0\ee^2 - \ee^2\ee^0)\zeta\n
&=& \frac{1}{4}(4\ee^1\ee^2 - 4\ee^2\ee^1 + 4i\ee^0\ee^3 - 4i\ee^3\ee^0)\zeta\n
&=& \frac{1}{4}(2i\ee^0\w\ee^3 - 2\ee^1\w\ee^2)\zeta\n
&=& i \left(\frac{1}{2}(\ee^0\w\ee^3 + i \ee^1\w\ee^2)\right)\zeta\n
&=&  i \left(\frac{1}{2}(\ee_\zeta^0\vcz\w\ee_\zeta^3 + i \ee_\zeta^1\vcz\w\ee_\zeta^2)\right),\qquad \text{from Eq.(\ref{1006})}\n
&=&i \lambda_\zeta^3\eeq\noi
\med
More generally the generators  ${\lambda^a_\zeta}$ 
satisfy the properties $[{\lambda^a_\zeta},{\lambda^b_\zeta}]_{\zeta} = if_{abc}\dl^{-1/2}\lambda^c_\zeta,$ 
 where the $\mk{su}(3)$ structure constants $f_{abc}$ are given by 
$f_{123} = 2f_{147} = -2f_{156} = 2f_{246} = 2f_{257} = 2f_{345} = -2f_{367} = 2f_{458}/\sqrt{3} = 2f_{678}/\sqrt{3} = 1$ \cite{prb}

It is also immediate to note that the elements $\{\lambda_\zeta^1, \lambda_\zeta^2, \lambda_\zeta^3\}$ and
 $\lambda_\zeta^8$ generates the isotopic subalgebra  
$\mk{su}(2)_\zeta\times\mk{u}(1)_\zeta$. 
\subsubsection*{SU(3)$_\zeta$: case 2}
By isotopically lifting the Lie algebra $\mk{su}(3)$ presented in \cite{romo,chi1} above it follows that
\beq
\xi_\zeta^1 &=& -\frac{i}{2}(\ee_\zeta^2\vcz\w\ee_\zeta^3 + i\ee_\zeta^0\vcz\w\ee_\zeta^2\vcz\w\ee_\zeta^3),\qquad \xi_\zeta^2 = \frac{i}{2}(\ee_\zeta^1\vcz\w\ee_\zeta^3 + \ee_\zeta^0\vcz\w\ee_\zeta^1\vcz\w\ee_\zeta^3),\n
\xi_\zeta^3 &=& \frac{i}{2}(\ee_\zeta^1\vcz\w\ee_\zeta^2 + \ee_\zeta^0\vcz\w\ee_\zeta^1\vcz\w\ee_\zeta^2),\qquad \xi_\zeta^4 = \me(\ee^5_\zeta + i\ee_\zeta^0\vcz\w\ee_\zeta^3),\n
\xi_\zeta^5 &=& \me(\ee_\zeta^3 - i\ee_\zeta^1\vcz\w\ee_\zeta^2\vcz\w\ee_\zeta^3),\qquad \xi_\zeta^6 = \me(\ee_\zeta^2 + i\ee_\zeta^0\vcz\w\ee_\zeta^1),\n
\xi_\zeta^7 &=& \me (\ee_\zeta^1 -i\ee_\zeta^0\vcz\w\ee_\zeta^2),\qquad \xi_\zeta^8 = \frac{i}{2\sqrt{3}}(2\ee_\zeta^0 + i\ee_\zeta^1\vcz\w\ee_\zeta^2 - i\ee_\zeta^0\vcz\w\ee_\zeta^1\vcz\w\ee_\zeta^2).
\eeq
From the generators $\{\xi_\zeta^a,\lambda_\zeta^a\}\equiv\mho^a$ of the isotopic Lie algebra  $\mk{su}(3)_\zeta$, 
the isotopic Lie group SU(3)$_\zeta$ is constructed  
via the \emph{iso}exponenciation defined by
\bege
\exp(\vcz\theta_a\di\mho^a), \quad \vcz\theta_a \in \vcz\CC.
\enge

\section{Isotopic lifting SU$_\zeta$($n$) of SU($n$)}
\label{sun}
It has been stated in \cite{coun} that all the elements of the compact spin groups Spin($n$,0) $\simeq$ Spin(0,$n$) 
are exponentials of bivectors when $n > 1$. Also, the same holds for the other spin groups only for Spin$_+$($n-1$,1) $\simeq$
 Spin$_+$(1,$n-1$), $n > 4$ \cite{coun}.
It is well-known that at least the Lie algebras of type $\mk{spin}(p,q)\simeq\mk{so}(p,q)$ associated with the spin groups above 
can be described as elements of a Clifford algebra endowed with the commutator
$(\clp, [\;\;,\;\;])$ for $p+q = n$ big enough \cite{sobcik}. 
Such statement for the Lie algebra $\mathfrak{so}(p,q)\simeq(\la^2(\RR^{p,q}),[\;\;,\;\;])$ associated with the group  SO($p,q$) 
can be immediately proved. 
The group SU($n$) can be constructed in the context of the Clifford algebra
 $(\cl_{2(p+q)},[\;\,,\;\,])$, by taking a basis $\{\ee_a\}_{a=1}^{2n}$ of $\RR^{p,q}$, 
where $\ee_a^2 = 1$ and $p+q=n$, and defining the 2-forms \cite{romo}
\beq
E^{pq} &=& \ee^p\w \ee^q + \ee^{p+n}\w \ee^{q+n}\n
F^{pq} &=& \ee^p\w\ee^{q+n} - \ee^{p+n}\w\ee^{q}\n
H^r    &=& \ee^r\w\ee^{r+n} -\ee^{r+n+1}\w\ee^{r+n+1}
\eeq\noi 
for $p,q = 1,\ldots,n$, $p\neq q$ and $k=1,\ldots, n-1$. It follows the expressions
\beq\label{01}
[E^{pq}, E^{st}] = 2E^{qt},\quad  [E^{pq}, F^{pq}] = -2H^q,\quad [E^{pq},E^{st}] = 0,\quad  [H^q, H^p] = 0,
\eeq\noi and also \beq\label{02}
[F^{pq}, F^{ps}] = 2E^{qs},\quad  [H^p, E^{pq}] = -2F^{pq}, \quad[F^{pq}, F^{st}] = 0, \quad [H^p, E^{qs}] = 2F^{qs},
\eeq\noi where the last commutator is non-trivial only when $q = p+1$. Relations given by Eqs.(\ref{01},\ref{02}) 
completely define the Lie algebra $\mk{su}(n)$ associated with the group SU($n$) \cite{romo,sobcik}. 

Now the generators of the isotopic group SU$_\zeta$($n$) are written using the isotopic exterior product defined by
 Eq.(\ref{zet1}), denoting $\ee^m_\zeta = \vcz{\ee^m}$, as
\beq
E_\zeta^{pq} &=& \ee_\zeta^p\vcz\w \ee_\zeta^q + \ee_\zeta^{p+n}\vcz\w \ee_\zeta^{q+n}\n
F_\zeta^{pq} &=& \ee_\zeta^p\vcz\w\ee_\zeta^{q+n} - \ee_\zeta^{p+n}\vcz\w\ee_\zeta^{q}\n
H_\zeta^r    &=& \ee_\zeta^r\vcz\w\ee_\zeta^{r+n} -\ee_\zeta^{r+n+1}\vcz\w\ee_\zeta^{r+n+1},
\eeq\noi where $p,q = 1,\ldots,n$, $p\neq q$ and $r=1,\ldots, n-1$. It follows that the expressions
\beq
[E_\zeta^{pq}, E_\zeta^{st}]_\zeta = 2E_\zeta^{qt},\quad  [E_\zeta^{pq}, F_\zeta^{pq}]_\zeta = -2H_\zeta^q,\quad [E_\zeta^{pq},E_\zeta^{st}]_\zeta = 0,\quad  [H_\zeta^q, H_\zeta^p]_\zeta = 0,
\eeq\noi and the relations \beq
[F_\zeta^{pq}, F_\zeta^{ps}]_\zeta = 2E_\zeta^{qs},\quad  [H_\zeta^p, E_\zeta^{pq}]_\zeta = -2F_\zeta^{pq}, \quad[F_\zeta^{pq}, F_\zeta^{st}]_\zeta = 0, \quad [H_\zeta^p, E_\zeta^{qs}]_\zeta
 = 2F_\zeta^{qs},\label{iso100}
\eeq\noi completely define  SU$_\zeta$($n$).

\section{Applications in flavor SU($n$) group symmetry}
\label{tro1}

Consider a Hilbert space ${\mt H}$ --- and ideal with respect to the algebra defined by the operators acting on it --- 
with elements $\{|\psi_i\rangle, \ldots\}$, where $\langle\psi_i|\psi_j\rangle\in\CC$,  
and the normalized states are given by $\langle\psi_i|\psi_j\rangle = \delta_{ij}$.  
In order to formulate the isotopic quantum mechanics, denominated relativistic hadronic mechanics, 
(RHM) \cite{san05}, 
consider now a Hilbert \emph{iso}space ${\mt H}_\zeta$, which has operators acting on its elements 
satisfying the the rule given by Eq.(\ref{78}). Elements $|\vcz\psi\rangle\in{\mt H}_\zeta$, and elements of the dual space 
$\langle\vcz\psi|\in{\mt H}_\zeta^*$ satisfy
\bege
\langle\vcz{\psi}\wr\vcz{\phi}\rangle:= \langle\vcz{\psi}| \zeta^{-1}|\vcz{\phi}\rangle \zeta\in\vcz{\CC}.
\enge\noi  In this case the normalized states are given by 
$\langle\vcz{\psi}\wr\vcz{\phi}\rangle = \zeta \in\vcz{\CC}$. 
With these definitions, Santilli shows that Hermitean (observable) 
operators in the quantum mechanical formalism correspond to isohermitean states in RHM.  

Hereon $\zeta$-kets $\wr\;\cdot\;\rangle$ are introduced
\bege
{}{\wr\psi\rangle:= \zeta^{-1}|\psi\rangle}
\enge
\noi together with the eigenvalue isoequation given by 
\bege
\vcz{H}\di |\vcz{\psi}\rangle = \vcz{H}\zeta^{-1}|\vcz{\psi}\rangle = \vcz{E} \di|\vcz{\psi}\rangle = E|\vcz{\psi}\rangle,
\enge\noi where $|\vcz\psi\rangle$ is an element of ${\mt H}_\zeta$ and  $H$ denotes an arbitrary operator 
acting on ${\mt H}_\zeta$.

\subsection{Exact flavor SU(3) symmetry, isomesons and isobarions}
\label{su3ex}
Hereon the formalism used is implicitly the Clifford algebra $\cle(\CC)$, since SU(3) is described in terms of $\cle(\CC)$, as in Subsec. (\ref{su3c}).
In Subsecs. (\ref{su3ex}) and (\ref{su6ex}) we use the most recent limits of quark masses given by \cite{prb}, page 36:
\beq
1.5\;\; {\rm MeV} &\leq m_u \leq& 3.0 \;\;{\rm MeV},\qquad 3\;\; {\rm MeV}\leq m_d\leq 7\;\; {\rm MeV},\n 70\;\;{\rm MeV}&\leq  m_s \leq& 110\;\; {\rm MeV},
\qquad
1.16\;\; {\rm GeV}\leq m_c \leq 1.34\;\; {\rm GeV},\n 4.13\;\;{\rm GeV}&\leq m_b\leq& 4.27\;\; {\rm GeV},\qquad 170.9\;\;{\rm GeV} \leq m_t \leq
 177.5\;\; {\rm GeV}.\label{iso0}
\eeq \noi in order to determine the isotopic element $\zeta$, which is shown to be 
function of these masses. In this sense quark masses are responsible 
for the deformation of the algebraic structures involved, together with the induced deformation concerning the
 geometric structure associated with the formalism presented here.

In this Section we briefly recall the lifting of SU(3) \cite{san22,san18,animalu}, in the
context introduced in this paper, which main aim is to extend the method
to an exact symmetry in the isotopic lifting of SU(6).
We must emphasize that since SU(3) $\subset {\mt M}(3,\CC)$, and ${\mt M}(4,\CC)\simeq\cle(\CC)$, the characterization of SU(3) in $\cle(\CC)$ can be 
done by considering the trivial `block' immersion of elements of ${\mt M}(3,\CC)$ in ${\mt M}(4,\CC)$, as
$A\mapsto {\fn{\bx A& \vec{0}\\\vec{0}^T&1\ex}}\in{\mt M}(4,\CC)$, where $A\in{\mt M}(3,\CC)$, $\vec{0} = (0,0,0)^T$ and 1 $\in \RR$. 

Considering a basis  $\{|\psi_u\rangle, |\psi_d\rangle, |\psi_s\rangle\}$ of the carrier representation space of SU(3), 
in a Hilbert space ${\mt H}$,  we now extend Santilli's  idea \cite{san18} describing the flavor SU(3) symmetry among quarks $u, d$ and $s$, 
in such a way that they consequently have the same mass, in isospace.

If we choose the representation of the isounit in ${\mt M}(3,\CC)\hko{\mt M}(4,\CC)\simeq\cle(\CC)$, as being 
 ${\zeta}$ = diag($g_{11}, g_{22}, g_{33}$,1), the Gell-Mann isomatrices are simply representations 
of the elements introduced in Eqs.(\ref{51}), when the Weyl (chiral) representation of $\{\ee_\mu\}$ is considered:
\beq
{\lam_1^\zeta} &=& \delta^{-1/2}\bx 0&g_{11}&0\\g_{22}&0&0\\0&0&0\ex,\qquad
{\lam_2^\zeta} = \dl^{-1/2} \begin{pmatrix}0&-ig_{11}&0\\ig_{22}&0&0\\0&0&0\ex,\n
{\lam_3^\zeta} &=& \dl^{-1/2}\bx g_{11}&0&0\\0&-g_{22}&0\\0&0&0\ex,\qquad
{\lam_4^\zeta} = \dl^{-1/2} \begin{pmatrix}0&0&g_{11}\\0&0&0\\g_{33}&0&0\ex,\n 
{\lam_5^\zeta} &=& \dl^{-1/2}\bx 0&0&-ig_{11}\\0&0&0\\ig_{33}&0&0\ex, \qquad
{\lam_6^\zeta} = \dl^{-1/2}\begin{pmatrix}0&0&0\\0&0&g_{22}\\0&g_{33}&0\ex,\n 
{\lam_7^\zeta} &=& \dl^{-1/2} \bx 0&0&0\\0&0&-ig_{22}\\0&ig_{33}&0\ex,\qquad
{\lam_8^\zeta} = \frac{\dl^{-1/2}}{\sqrt{3}}\bx g_{11}&0&0\\0&g_{22}&0\\0&0&-2g_{33}\ex,
\eeq
\noi and satisfy the properties $[{\lambda_a^\zeta},{\lambda_b^\zeta}]_{\zeta} = 2if_{abc}\dl^{-1/2}\lambda_c^\zeta,$ 
 where the $\mk{su}(3)$ structure constants $f_{abc}$ are given by 
$f_{123} = 2f_{147} = -2f_{156} = 2f_{246} = 2f_{257} = 2f_{345} = -2f_{367} = 2f_{458}/\sqrt{3} = 2f_{678}/\sqrt{3} = 1$ \cite{prb}.

With the condition $\det{\zeta} = 1$, we endow the 
 Gell-Mann isomatrices with a standard adjoint representation character \cite{san12,san18}.
Such condition implies that $\zeta$ can be written as  
\bege
{\bf \zeta} = {\rm diag}(\alpha^{-1}, \be^{-1}, \alpha\be,1), \qquad \alpha,\be\in\RR.
\enge\noi
Using this choice the isonormalized isostates are given by 
\bege
|\vcz{\psi}_u\rangle =  \left(\bea{c}\alpha^{-1/2}\\ 0\\ 0\ear\right),\qquad |\vcz{\psi}_d\rangle = \left(
\bea{c}0\\ \be^{-1/2}\\ 0\ear\right),\qquad |\vcz{\psi}_s\rangle =  \left(\bea{c} 0\\ 0\\
 (\alpha\be)^{1/2}\ear\right)\enge\noi and satisfy the relations $\langle\vcz{\psi_i}\wr\vcz{\psi_j}\rangle = \delta_{ij}\zeta$. 

Since the mass operator in $\mk{su}(3)$ is given by  
\beq
M &=& \frac{1}{3}(m_u + m_d + m_s) I + \me(m_u - m_d)\lam^3 + \frac{\sqrt{3}}{6}(m_u + m_d - 2m_s)\lam^8\n
&=& {\rm diag}\, (m_u, m_d, m_s),
\eeq\noi the isotopic lifting SU(3)$_\zeta$ of SU(3) has the mass operator given by 
\beq
\vcz{M} &=& \left(\frac{1}{3}(m_u + m_d + m_s) {\zeta} + \me(m_u - m_d){\lam_\zeta^3} + \frac{\sqrt{3}}{6}(m_u + m_d - 2m_s){\lam_\zeta^8}\right){\zeta}\n
&=& {\rm diag}\, (\alpha^{-1}m_u, \be^{-1}m_d, \alpha\be m_s).
\eeq\noi In the simultaneous limits $\alpha\Ra 1$ and  $\be\Ra 1$, it can be verified that $\vcz{M}\Ra M$.  
We now constrain the parameters $\alpha,\be$ that compose the isounit $\zeta$, imposing that in isospace quarks $u,d$ and $s$ have the \emph{same} mass 
 $\vcz{m} =   \alpha^{-1}m_\zeta = \be^{-1}m_d = \alpha\be m_s$. Then,  and $\alpha,\be$ are shown to be functions of quarks $u,d$, $s$ masses, given explicitly by 
\bege
\alpha = \left(\frac{m_u^2}{m_s m_d}\right)^{1/3},\qquad  \be = \left(\frac{m_d^2}{m_s m_u}\right)^{1/3}.
\enge\noi 
Taking the masses values in Eq.(\ref{iso0}), the most recent limits of  $\alpha$ and  $\be$ are given by 
\bege
{}{0.2204 \leq \alpha \leq 0.2638\qquad 0,2768 \leq \be \leq 0.3057}
\enge\noi The exactness, or a better approximation for the values of $\alpha$ and $\be$ relies on the precision in the determination of the
masses $m_u, m_d$ and $m_s$. 

Although the masses $m_u, m_d$ and $m_s$ are imposed to be equal in isospace, in physical space the conventional values 
of quarks $u, d$ and $s$ masses are given by the eigenvalue isoequation 
\bege
\vcz{M}\di|\vcz{\psi}\rangle = M\zeta \zeta^{-1}|\vcz{\psi}\rangle = M|\vcz{\psi}\rangle =  {\rm diag}\,(m_u, m_d, m_s) |\vcz{\psi}\rangle
\enge\noi or, equivalently, via expected values:
\bege\langle\vcz{\psi_u}\wr \vcz{M}\wr\vcz{\psi_u}\rangle = m_u,\qquad \langle\vcz{\psi_d}\wr \vcz{M}\wr\vcz{\psi_d}\rangle = m_d, 
\qquad \langle\vcz{\psi_s}\wr \vcz{M}\wr\vcz{\psi_s}\rangle = m_s.
\enge\noi The hypercharge operator  $Y$ is naturally extended to isospace as 
\bege
\vcz{Y} = \frac{1}{2\sqrt{3}}\vcz{\lam_8} = \frac{1}{2\sqrt{3}}\,{\rm diag}\,(\alpha^{-1}, \be^{-1}, -2(\alpha\be)),
\enge\noi while the  $z$ isospin component $I_3$ is given by
\bege
\me\vcz{\lam_3} = {\rm diag}\,(\alpha, -\be, 0).\enge
Indeed, the expected eigenvalues for the operators  above are
\beq
Y(u) &=& \langle\vcz{\psi_u}\wr {Y}\wr\vcz{\psi_u}\rangle = \frac{1}{6},\qquad Y(d) = \langle\vcz{\psi_d}\wr {Y}\wr\vcz{\psi_d}\rangle = \frac{1}{6},\n 
Y(s) &=& \langle\vcz{\psi_s}\wr {Y}\wr\vcz{\psi_s}\rangle = -\frac{1}{3} 
\eeq\noi and
\beq
I_3(u) &=& \langle\vcz{\psi_u}\wr {I_3}\wr\vcz{\psi_u}\rangle = \frac{1}{2},\qquad I_3(d) = \langle\vcz{\psi_d}\wr {I_3}\wr\vcz{\psi_d}\rangle = -\frac{1}{2},\n 
I_3(s) &=& \langle\vcz{\psi_s}\wr {I_3}\wr\vcz{\psi_s}\rangle = 0. 
\eeq\noi The isotopic electric charge operator is obviously given by $
Q = {Y} + {I_3}$.

Now denoting $|\psi\rangle$ a state describing any of the quarks $\{|\psi_u,\psi_d,\psi_s\}$, an isotopic lifting 
 induces  mesons, described by $|\psi\rangle\ot\bar{|\psi\rangle}$, and barions, described by $|\psi\rangle\ot|\psi\rangle\ot|\psi\rangle$,  
to have the corresponding states in isospace
\bege
|\vcz\psi\rangle\ot_\zeta|{\vcz{\overline\psi}}\rangle
\enge\noi for the mesons
\bege
|\vcz\psi\rangle\ot_\zeta|\vcz\psi\rangle\ot_\zeta|\vcz\psi\rangle,
\enge\noi for the barions. The symbol $\ot_\zeta$ denotes the \emph{iso}tensorial product between spinor fields in $\CC\ot\cle$ by 
\bege
{(\;{\cdot}\;)\;\ot_\zeta\;(\;{\cdot}\;):={{}{\zeta}}^{-1}(\;\cdot\;)\;\ot\;(\;\cdot\;)(\widetilde{{{}{\zeta}}^{-1}\ee_0})^*}\enge\noi
The isomeson can be expressed as 
\bege {\text{isomeson}}\;=\;{{}{\zeta}}^{-1}\;\;{\text{meson}}\;\;(\widetilde{{{\zeta}}^{-1}\ee_0})^*\enge

\subsection{Exact flavor SU(6) isotopic symmetry}
\label{su6ex}
Up to now, there is no description of the generators of the group SU(6) in terms of elements of any of the minimal Clifford algebras
$\cl_{1,6}\simeq \cl_{3,4}\simeq \cl_{5,2}\simeq {\mt M}(8,\CC)$.  Up to our knowledge, there is not any explicit construction 
like in Section (IX), where in \cite{romo} SU(3) is constructed inside the Dirac algebra $\cle$, and here we have extended it 
to the isotopic case. Although there is not such an explicit construction, it is still possible to consider the isotopic lifting 
of the generators of the representation of SU(6) --- seen as a subgroup of 
${\mt M}(6,\CC) \hookrightarrow {\mt M}(8,\CC)\simeq\cl_{1,6}\simeq \cl_{3,4}\simeq \cl_{5,2}$. Also, 
 the characterization of SU(6) in ${\mt M}(8,\CC)$ can be accomplished
if the trivial `block' immersion of elements of SU(6)$\hookrightarrow{\mt M}(6,\CC)$ in ${\mt M}(8,\CC)$, as
\bege
\label{inc}B\mapsto {\fn{\bx B& \vec{0}&\vec{0}\\\vec{0}^T&1&0\\\vec{0}^T&0&1\ex}}\in{\mt M}(8,\CC),
\enge\noi where $B\in {\rm SU(6)}\hookrightarrow 
{\mt M}(6,\CC)$, $\vec{0} = (0,0,0,0,0,0)^T$, $0\in \CC$, and 1 $\in \RR$. 

A particular case of Eqs.(\ref{iso100}), relating exterior algebra elements in $\RR^{12,0}$ and the generators of $\mathfrak{su}(6)$ 
is considered in details in the Appendix.

Now a basis  $\{|\psi_u\rangle, |\psi_d\rangle, |\psi_s\rangle, |\psi_c\rangle, |\psi_b\rangle, |\psi_t\rangle \}$ 
of the carrier representation space of SU(6) 
in a Hilbert space ${\mt H}$ is considered,  an Santilli's  idea \cite{san22,san18,animalu} is extended to 
describe the flavor SU(6) symmetry among quarks $u, d$, $s$ $c$, $b$, and $t$, 
in such a way that they consequently have the same mass, in isospace.

If we choose the representation of the isounit in SU(6)$\hko{\mt M}(6,\CC)\hko{\mt M}(8,\CC)\simeq\cle(\CC)$, as being 
 ${\zeta}$ = diag($g_{11}, g_{22}, g_{33}, g_{44}, g_{55}$,1), we can describe SU(6) generators by
\beq
{\lam_1^\zeta} &=& \delta^{-1/2}\bx 0&g_{11}&0&0&0&0\\g_{22}&0&0&0&0&0\\0&0&0&0&0&0\\0&0&0&0&0&0\\0&0&0&0&0&0\\0&0&0&0&0&0\ex,\qquad
{\lam_2^\zeta} = \dl^{-1/2} \begin{pmatrix}0&-ig_{11}&0&0&0&0\\ig_{22}&0&0&0&0&0\\0&0&0&0&0&0\\0&0&0&0&0&0\\0&0&0&0&0&0\\0&0&0&0&0&0\ex,\n
{\lam_3^\zeta} &=& \dl^{-1/2}\bx g_{11}&0&0&0&0&0\\0&-g_{22}&0&0&0&0\\0&0&0&0&0&0\\0&0&0&0&0&0\\0&0&0&0&0&0\\0&0&0&0&0&0\ex,\qquad
{\lam_4^\zeta} = \dl^{-1/2} \begin{pmatrix}0&0&g_{11}&0&0&0\\0&0&0&0&0&0\\g_{33}&0&0&0&0&0\\0&0&0&0&0&0\\0&0&0&0&0&0\\0&0&0&0&0&0\ex,\n 
{\lam_5^\zeta} &=& \dl^{-1/2}\bx 0&0&-ig_{11}&0&0&0\\0&0&0&0&0&0\\ig_{33}&0&0&0&0&0\\0&0&0&0&0&0\\0&0&0&0&0&0\\0&0&0&0&0&0\ex, \qquad
{\lam_6^\zeta} = \dl^{-1/2}\begin{pmatrix}0&0&0&0&0&0\\0&0&g_{22}&0&0&0\\0&g_{33}&0&0&0&0\\0&0&0&0&0&0\\0&0&0&0&0&0\\0&0&0&0&0&0\ex,\n 
{\lam_7^\zeta} &=& \dl^{-1/2} \bx 0&0&0&0&0&0\\0&0&-ig_{22}&0&0&0\\0&ig_{33}&0&0&0&0\\0&0&0&0&0&0\\0&0&0&0&0&0\\0&0&0&0&0&0\ex,\qquad
{\lam_8^\zeta} = \frac{\dl^{-1/2}}{\sqrt{3}}\bx g_{11}&0&0&0&0&0\\0&g_{22}&0&0&0&0\\0&0&-2g_{33}&0&0&0\\0&0&0&0&0&0\\0&0&0&0&0&0\\0&0&0&0&0&0\ex,\eeq\beq
{\lam_9^\zeta} &=& \delta^{-1/2}\bx 0&0&0&g_{11}&0&0\\0&0&0&0&0&0\\0&0&0&0&0&0\\g_{44}&0&0&0&0&0\\0&0&0&0&0&0\\0&0&0&0&0&0\ex,\qquad
{\lam_{10}^\zeta} = \delta^{-1/2}\bx 0&0&0&-ig_{11}&0&0\\0&0&0&0&0&0\\0&0&0&0&0&0\\ig_{44}&0&0&0&0&0\\0&0&0&0&0&0\\0&0&0&0&0&0\ex,\n
{\lam_{11}^\zeta} &=& \dl^{-1/2}\bx 0&0&0&0&0&0\\0&0&0&g_{22}&0&0\\0&0&0&0&0&0\\0&g_{44}&0&0&0&0\\0&0&0&0&0&0\\0&0&0&0&0&0\ex,\qquad
{\lam_{12}^\zeta} = \dl^{-1/2} \bx 0&0&0&0&0&0\\0&0&0&-ig_{22}&0&0\\0&0&0&0&0&0\\0&ig_{44}&0&0&0&0\\0&0&0&0&0&0\\0&0&0&0&0&0\ex,\n
{\lam_{13}^\zeta} &=& \dl^{-1/2}\bx 0&0&0&0&0&0\\0&0&0&0&0&0\\0&0&0&g_{33}&0&0\\0&0&g_{44}&0&0&0\\0&0&0&0&0&0\\0&0&0&0&0&0\ex, \qquad
{\lam_{14}^\zeta} = \dl^{-1/2}\bx 0&0&0&0&0&0\\0&0&0&0&0&0\\0&0&0&-ig_{33}&0&0\\0&0&ig_{44}&0&0&0\\0&0&0&0&0&0\\0&0&0&0&0&0\ex, \n
{\lam_{15}^\zeta} &=& \frac{\dl^{-1/2}}{\sqrt{6}}\bx g_{11}&0&0&0&0&0\\0&g_{22}&0&0&0&0\\0&0&g_{33}&0&0&0\\0&0&0&-3g_{44}&0&0\\0&0&0&0&0&0\\0&0&0&0&0&0\ex,\n
{\lam_{16}^\zeta} &=& \delta^{-1/2}\bx 0&0&0&0&g_{11}&0\\0&0&0&0&0&0\\0&0&0&0&0&0\\0&0&0&0&0&0\\g_{55}&0&0&0&0&0\\0&0&0&0&0&0\ex,\qquad
{\lam_{17}^\zeta} = \delta^{-1/2}\bx 0&0&0&0&-ig_{11}&0\\0&0&0&0&0&0\\0&0&0&0&0&0\\0&0&0&0&0&0\\ig_{55}&0&0&0&0&0\\0&0&0&0&0&0\ex,\n
{\lam_{18}^\zeta} &=& \dl^{-1/2}\bx 0&0&0&0&0&0\\0&0&0&0&g_{22}&0\\0&0&0&0&0&0\\0&0&0&0&0&0\\0&g_{55}&0&0&0&0\\0&0&0&0&0&0\ex,\qquad
{\lam_{19}^\zeta} = \dl^{-1/2} \bx 0&0&0&0&0&0\\0&0&0&0&-ig_{22}&0\\0&0&0&0&0&0\\0&0&0&0&0&0\\0&ig_{55}&0&0&0&0\\0&0&0&0&0&0\ex,\n
{\lam_{20}^\zeta} &=& \dl^{-1/2}\bx 0&0&0&0&0&0\\0&0&0&0&0&0\\0&0&0&0&g_{33}&0\\0&0&0&0&0&0\\0&0&g_{55}&0&0&0\\0&0&0&0&0&0\ex, \qquad
{\lam_{21}^\zeta} = \dl^{-1/2}\bx 0&0&0&0&0&0\\0&0&0&0&0&0\\0&0&0&0&-ig_{33}&0\\0&0&0&0&0&0\\0&0&ig_{55}&0&0&0\\0&0&0&0&0&0\ex, \label{g81}
\eeq\noi 
\beq
{\lam_{22}^\zeta} &=& \dl^{-1/2}\bx 0&0&0&0&0&0\\0&0&0&0&0&0\\0&0&0&0&0&0\\0&0&0&0&g_{44}&0\\0&0&0&g_{55}&0&0\\0&0&0&0&0&0\ex, \qquad
{\lam_{23}^\zeta} = \dl^{-1/2}\bx 0&0&0&0&0&0\\0&0&0&0&0&0\\0&0&0&0&0&0\\0&0&0&0&-ig_{44}&0\\0&0&0&ig_{55}&0&0\\0&0&0&0&0&0\ex, \n
{\lam_{24}^\zeta} &=& \frac{\dl^{-1/2}}{2\sqrt{6}}\bx g_{11}&0&0&0&0&0\\0&g_{22}&0&0&0&0\\0&0&g_{33}&0&0&0\\0&0&0&g_{44}&0&0
\\0&0&0&0&-4g_{55}&0\\0&0&0&0&0&0\ex,\qquad
{\lam_{25}^\zeta} = \delta^{-1/2}\bx 0&0&0&0&0&g_{11}\\0&0&0&0&0&0\\0&0&0&0&0&0\\0&0&0&0&0&0\\0&0&0&0&0&0\\g_{66}&0&0&0&0&0\ex,\n
{\lam_{26}^\zeta} &=& \delta^{-1/2}\bx 0&0&0&0&0&-ig_{11}\\0&0&0&0&0&0\\0&0&0&0&0&0\\0&0&0&0&0&0\\0&0&0&0&0&0\\ig_{66}&0&0&0&0&0\ex,\qquad
{\lam_{27}^\zeta} = \dl^{-1/2}\bx 0&0&0&0&0&0\\0&0&0&0&0&g_{22}\\0&0&0&0&0&0\\0&0&0&0&0&0\\0&0&0&0&0&0\\0&g_{66}&0&0&0&0\ex,\n
{\lam_{28}^\zeta} &=& \dl^{-1/2} \bx 0&0&0&0&0&0\\0&0&0&0&0&-ig_{22}\\0&0&0&0&0&0\\0&0&0&0&0&0\\0&0&0&0&0&0\\0&ig_{66}&0&0&0&0\ex,\qquad
{\lam_{29}^\zeta} = \dl^{-1/2}\bx 0&0&0&0&0&0\\0&0&0&0&0&0\\0&0&0&0&0&g_{33}\\0&0&0&0&0&0\\0&0&0&0&0&0\\0&0&g_{66}&0&0&0\ex, \n
{\lam_{30}^\zeta} &=& \dl^{-1/2}\bx 0&0&0&0&0&0\\0&0&0&0&0&0\\0&0&0&0&0&-ig_{33}\\0&0&0&0&0&0\\0&0&0&0&0&0\\0&0&ig_{66}&0&0&0\ex, \qquad
{\lam_{31}^\zeta} = \dl^{-1/2}\bx 0&0&0&0&0&0\\0&0&0&0&0&0\\0&0&0&0&0&0\\0&0&0&0&0&g_{44}\\0&0&0&0&0&0\\0&0&0&g_{66}&0&0\ex, \n
{\lam_{32}^\zeta} &=& \dl^{-1/2}\bx 0&0&0&0&0&0\\0&0&0&0&0&0\\0&0&0&0&0&0\\0&0&0&0&0&-ig_{44}\\0&0&0&0&0&0\\0&0&0&ig_{66}&0&0\ex, \qquad
{\lam_{33}^\zeta} = \dl^{-1/2}\bx 0&0&0&0&0&0\\0&0&0&0&0&0\\0&0&0&0&0&0\\0&0&0&0&0&0\\0&0&0&0&0&g_{55}\\0&0&0&0&g_{66}&0\ex, \n
{\lam_{34}^\zeta} &=& \dl^{-1/2}\bx 0&0&0&0&0&0\\0&0&0&0&0&0\\0&0&0&0&0&0\\0&0&0&0&0&0\\0&0&0&0&0&-ig_{55}\\0&0&0&0&ig_{66}&0\ex, \qquad
{\lam_{35}^\zeta} = \frac{\dl^{-1/2}}{2\sqrt{30}}\bx g_{11}&0&0&0&0&0\\0&g_{22}&0&0&0&0\\0&0&g_{33}&0&0&0\\0&0&0&g_{44}&0&0
\\0&0&0&0&g_{55}&0\\0&0&0&0&0&-5g_{66}\ex,\label{g82}
\eeq

With the condition $\det{\zeta} = 1$, we endow the 
 Gell-Mann isomatrices with a standard adjoint representation character \cite{san12,san18}.
Such condition implies that $\zeta$ can be written as  
\bege
{\bf \zeta} = {\rm diag}(\alpha^{-1}, \be^{-1}, \omega^{-1}, \kappa^{-1}, \tau^{-1}, \alpha\be\omega\kappa\tau,1,1), \qquad \alpha,\be,\omega,\kappa,\tau \in\RR.
\enge\noi
Using this choice the isonormalized isostates are given by 
\beq
|\vcz{\psi}_u\rangle &=&  \left(\bea{c}\alpha^{-1/2}\\ 0\\ 0\\0\\0\\0\ear\right),\qquad |\vcz{\psi}_d\rangle = \left(
\bea{c}0\\ \be^{-1/2}\\ 0\\0\\0\\0\ear\right),\qquad |\vcz{\psi}_s\rangle =  \left(\bea{c} 0\\ 0\\
 \omega^{-1/2}\\0\\0\\0\ear\right),\n 
|\vcz{\psi}_c\rangle &=&  \left(\bea{c} 0\\ 0\\
 0\\\kappa^{-1/2}\\0\\0\ear\right),\qquad 
|\vcz{\psi}_b\rangle =  \left(\bea{c} 0\\ 0\\0\\0\\\tau^{-1}\\0\ear\right),\qquad 
|\vcz{\psi}_t\rangle =  \left(\bea{c} 0\\ 0\\0\\0\\0\\(\alpha\beta\omega\kappa\tau)^{1/2}\ear\right),
\eeq\noi and satisfy the relations $\langle\vcz{\psi_a}\wr\vcz{\psi_b}\rangle = \delta_{ab}\zeta$, where here $\delta_{ab}$ denotes 
the Kronecker delta. 

Since the mass operator in SU(6) is given by  
\beq
\hspace{-0.82cm}M &=& \frac{1}{3}(m_u + m_d + m_s +m_c +m_b +m_t) I + \frac{107}{144}(m_u - m_d)\lam_3 \n
&&- \frac{55\sqrt{3}}{144}(m_u + m_d - 2m_s)\lam_8  
- \frac{55\sqrt{6}}{144}(m_u + m_d + m_s -3m_c)\lam_{15}\n
&& -\frac{11\sqrt{6}}{24}(m_u + m_d + m_s + m_c -4m_b)\lam_{24} -\frac{\sqrt{30}}{6}(m_u + m_d + m_s + m_c + m_b -5m_t)\lam_{35} 
\n
&=& {\rm diag}\, (m_u, m_d, m_s,m_c,m_b,m_t),
\eeq\noi the isotopic lifting SU(6)$_\zeta$ of SU(6) has the mass operator given by 
\beq
\hspace{-0.82cm}\vcz{M} &=& \frac{1}{3}(m_u + m_d + m_s +m_c +m_b +m_t) \zeta +
 \frac{107}{144}(m_u - m_d)\lam_3^\zeta \zeta\n&&- \frac{55\sqrt{3}}{144}(m_u + m_d - 2m_s)\lam_8^\zeta \zeta 
- \frac{55\sqrt{6}}{144}(m_u + m_d + m_s -3m_c)\lam_{15}^\zeta\zeta\n
&& -\frac{11\sqrt{6}}{24}(m_u + m_d + m_s + m_c -4m_b)\lam_{24}^\zeta \zeta-\frac{\sqrt{30}}{6}(m_u + m_d + m_s + m_c + m_b -5m_t)\lam_{35}^
\zeta {\zeta}\n
&=& {\rm diag}\, (\alpha^{-1}m_u, \be^{-1}m_d, \omega^{-1}m_s, \kappa^{-1}m_c, \tau^{-1} m_b, \alpha\,\beta\,\omega\,\kappa\,\tau\, m_t).
\eeq\noi

By extending the process of the previous subsection to the SU(6) isotopic lifting describing an exact flavor symmetry, 
the mass operator in isospace, considering six quarks, is presented as 
\beq\label{is1}\vcz{M} = M\zeta  &=& {\rm diag}\, (\alpha^{-1}m_u, \be^{-1}m_d, \omega^{-1} m_s, \kappa^{-1}m_c, \tau^{-1}m_b, \alpha\be\omega\kappa\tau \;m_t)\\
&\equiv& {\rm diag}\,(\vcz{m},\vcz{m},\vcz{m},\vcz{m},\vcz{m},\vcz{m})\label{is2}
\eeq\noi where in this case the isounit is given by $\zeta =$ diag ($\alpha^{-1}, \be^{-1}, \omega^{-1}, \kappa^{-1}, \tau^{-1}, \alpha\be\omega\kappa\tau)$.  Imposing Eq.(\ref{is2}), we see that each term of the matrix in 
Eq.(\ref{is1}) must equal each other, i.e.,
\bege\label{is3}
\alpha^{-1}m_u= \be^{-1}m_d= \omega^{-1} m_s= \kappa^{-1}m_c= \tau^{-1}m_b= \alpha\be\omega\kappa\tau \;m_t
\enge
and in particular, let us isolate all the variables in terms of the variable $\alpha$:
\bege\label{is4}
\beta = \alpha\frac{m_d}{m_u},\;\;\omega = \alpha\frac{m_s}{m_u},\;\;\kappa = \alpha\frac{m_c}{m_u},\;\;\tau = \alpha\frac{m_b}{m_u}.
\enge\noi Substituting Eqs.(\ref{is4}) in the last of Eqs.(\ref{is3}) 
$(\alpha^{-1}m_u= \alpha\be\omega\kappa\tau \;m_t)$ yields
\bege
\alpha^{-1}m_u = \alpha^5\frac{m_d}{m_u^4}m_sm_cm_bm_t
\enge\noi implying that 
\bege{\alpha = \left(\frac{m_u^5}{m_dm_sm_cm_bm_t}\right)^{1/6}}
\enge
In the same way it can be shown that 
\bege{\be = \left(\frac{m_d^5}{m_um_sm_cm_bm_t}\right)^{1/6}},\qquad {\omega = \left(\frac{m_s^5}{m_dm_um_cm_bm_t}\right)^{1/6}}
\enge
\bege{\kappa = \left(\frac{m_c^5}{m_dm_sm_um_bm_t}\right)^{1/6}}
{\tau = \left(\frac{m_b^5}{m_dm_sm_cm_um_t}\right)^{1/6}}
\enge
Substituting the values of quarks masses \cite{prb} in (\ref{iso0}) yields
{{\beq
&&5.945\times 10^{-3}\;\leq\;\alpha  \;\leq\; 8.212\times 10^{-3} \n 
&&1.189\times 10^{-2}\;\leq\;\be  \;\leq\;1.920\times 10^{-2}\n
&&2.774\times 10^{-1}\;\leq\;\omega  \;\leq\;3.018\times 10^{-1}\n 
&&3.676\;\leq\;\kappa  \;\leq\;4.598\n
&&486.938\;\leq\;\tau  \;\leq\;677.379\nonumber\eeq
}}

\section{Concluding Remarks}
This paper presents for the first time the
isotopies of Clifford algebras with relevant applications in flavor symmetry of quarks.
We have formulated the isotopic liftings in the context of Clifford algebras and
highlighted the formal description concerning isotopies for non-associative general algebras.
The structure of the isoalgebra identifies the the mass matrices to a multiple of the identity operator.
The formalism used is solely based on Clifford algebras, the more natural formalism which is able to introduce isotopies of the exterior algebra. 
The flavor hadronic symmetry of the six $u,d,s,c,b,t$ quarks is shown to be exact if 
the isotopic group SU(6) is regarded. 
We have shown that the unit of the isotopic Clifford algebra
is a function of the six quark masses. It illustrates how phenomenological data concerning quark masses can constrain the geometry
of spacetime, where the limits constraining the parameters, that are entries of the 
representation of the isounit in the isotopic group SU(6),  are based on the most recent limits imposed on quark masses.
We assert that the formulation of other theories in isospace can bring a new class of solutions 
of open questions in theoretical physics. 

\section{Appendix: Clifford algebra generators of $\mk{su}(6)$}

In order to completely define  $\mk{su}_\zeta$(6) in terms of the exterior algebra of $\vcz\cl_{12,0}$, in the light of Eqs.(\ref{iso100}),
\beq
E_\zeta^{pq} &=& \ee_\zeta^p\vcz\w \ee_\zeta^q + \ee_\zeta^{p+n}\vcz\w \ee_\zeta^{q+n}\n
F_\zeta^{pq} &=& \ee_\zeta^p\vcz\w\ee_\zeta^{q+n} - \ee_\zeta^{p+n}\vcz\w\ee_\zeta^{q}\n
H_\zeta^r    &=& \ee_\zeta^r\vcz\w\ee_\zeta^{r+n} -\ee_\zeta^{r+n+1}\vcz\w\ee_\zeta^{r+n+1},
\eeq\noi where $p,q = 1,\ldots,6$, $p\neq q$ and $r=1,\ldots, 5$, let us identify 
$\{\lam_1^\zeta,\ldots,\lam_{15}^\zeta\} = \{E_\zeta^{pq}\}$, $\{\lam_{16}^\zeta,\ldots,\lam_{30}^\zeta\} = \{F_\zeta^{pq}\}$, and 
$\{\lam_{31}^\zeta,\ldots,\lam_{35}^\zeta\} = \{H_\zeta^1, H_\zeta^2, H_\zeta^3, H_\zeta^4, H_\zeta^5\}$. 
By this identification it can be verified that the set $\{\lam_{1}^\zeta,\ldots,\lam_{35}^\zeta\}$, explicitly constructed in Eqs.
(\ref{g81}), (\ref{g82})
 of generators of $\mathfrak{su}(6)$ satisfy
 \beq
[E_\zeta^{pq}, E_\zeta^{st}]_\zeta = 2E_\zeta^{qt},\quad  [E_\zeta^{pq}, F_\zeta^{pq}]_\zeta = -2H_\zeta^q,\quad [E_\zeta^{pq},E_\zeta^{st}]_\zeta = 0,\quad  [H_\zeta^q, H_\zeta^p]_\zeta = 0,
\eeq\noi and the relations \beq
[F_\zeta^{pq}, F_\zeta^{ps}]_\zeta = 2E_\zeta^{qs},\quad  [H_\zeta^p, E_\zeta^{pq}]_\zeta = -2F_\zeta^{pq}, \quad[F_\zeta^{pq}, F_\zeta^{st}]_\zeta = 0, \quad [H_\zeta^p, E_\zeta^{qs}]_\zeta
 = 2F_\zeta^{qs},
\eeq\noi for $p,q = 1,\ldots,6$, $p\neq q$ and $r=1,\ldots, 5$.

It is well known that $\cl_{12,0}\simeq {\mt M}(32,\HH)$, and we included 
$SU(6)\hko {\mt M}(6,\CC)\hko{\mt M}(8,\CC)\hko{\mt M}(32,\HH)$, via Eq.(\ref{inc}) \emph{and} the inclusion 
$A \hko\bx A&0_8&0_8&0_8\\0_8&0_8&0_8&0_8\\0_8&0_8&0_8&0_8\\0_8&0_8&0_8&0_8\ex$, where $A\in {\mt M}(8,\CC)$, and  
$0_8 \equiv  0_{8\times 8} \in {\mt M}(8,\CC)$. Up to our knowledge, there is not any criterion or method to \emph{explicitly} include
$\mk{su}(n)$ in any Clifford algebra $\cl_{j}$, where $j<2n$.

\section*{Acknowledgments}

R. da Rocha thanks to Funda\c c\~ao de Amparo \`a Pesquisa
do Estado de S\~ao Paulo (FAPESP) for financial support.

\end{document}